\documentclass[a4paper]{article}

\usepackage{authblk}
\usepackage[pdftex]{graphicx,color}
\usepackage{lmodern}
\usepackage[T1]{fontenc}
\usepackage{amsthm}
\usepackage{amsmath}
\usepackage{amssymb}
\usepackage[breaklinks=true]{hyperref}
\usepackage{listings}
\usepackage[scaled=0.80]{beramono}
\usepackage{textcomp}

\newcommand{\comment}[1]{}
\newcommand{\mail}[1]{\href{mailto:#1}{\nolinkurl{#1}}}

\newcommand{\us}{\_}
\newcommand{\type}[1]{\texttt{#1}}
\newcommand{\class}[1]{\texttt{#1}}
\newcommand{\locale}[1]{\texttt{#1}}
\newcommand{\const}[1]{\textup{\texttt{#1}}}

\newcommand{\thm}[1]{\textit{#1}}
\newcommand{\thy}[1]{\textit{#1}}
\newcommand{\param}[1]{\underline{\textsl{\textsf{#1}}}}

\newcommand{\C}{\const{coeff}}
\newcommand{\ordp}{\preceq_{\sf{p}}}
\newcommand{\ordsp}{\prec_{\sf{p}}}
\newcommand{\ordv}{\preceq_{\sf{t}}}
\newcommand{\ordsv}{\prec_{\sf{t}}}
\newcommand{\dvdv}{\const{dvd}_{\sf{t}}}

\newcommand{\spoly}{\const{spoly}}
\newcommand{\lt}{\const{lt}}
\newcommand{\lp}{\const{lp}}
\newcommand{\lc}{\const{lc}}

\newcommand{\polym}[2]{#1\Rightarrow_0#2}
\newcommand{\polyab}{\polym{\alpha}{\beta}}
\newcommand{\polyakb}{\polym{(\alpha\times\kappa)}{\beta}}

\title{Gr\"obner Bases of Modules and\\Faug\`ere's $F_4$ Algorithm in 
Isabelle/HOL\\[\smallskipamount]\large -- extended version --}

\author[1]{
Alexander Maletzky\thanks{The research was funded by the Austrian 
Science Fund (FWF): P 29498-N31}
}
\author[2]{
Fabian Immler\thanks{The research was funded by DFG Koselleck 
Grant NI 491/16-1}
}
\affil[1]{
RISC, Johannes Kepler Universit\"at Linz, Austria,\newline
\texttt{alexander.maletzky@risc.jku.at}
}
\affil[2]{
Institut f\"ur Informatik, Technische Universit\"at M\"unchen, Germany,\newline
\texttt{immler@in.tum.de}
}

\date{}

\newtheorem{theorem}{Theorem}
\theoremstyle{definition}
\newtheorem{example}{Example}
\newtheorem{definition}[theorem]{Definition}
\theoremstyle{remark}
\newtheorem{remark}{Remark}

\begin{document}
\maketitle

\begin{abstract}
We present an elegant, generic and extensive formalization of Gr\"obner bases in Isabelle/HOL. The 
formalization covers all of the essentials of the theory (polynomial reduction, S-polynomials, 
Buchberger's algorithm, Buchberger's criteria for avoiding useless pairs), but also includes more 
advanced features like reduced Gr\"obner bases. Particular highlights are the first-time 
formalization of Faug\`ere's matrix-based $F_4$ algorithm and the fact that the entire theory is 
formulated for modules and submodules rather than rings and ideals. All formalized algorithms can be 
translated into executable code operating on concrete data structures, enabling the certified 
computation of (reduced) Gr\"obner bases and syzygy modules.
\end{abstract}


\section{Introduction}
\label{sec::Introduction}

Since their origins in Buchberger's PhD thesis~\cite{Buchberger65}, \emph{Gr\"obner bases} have 
become one of the most powerful and most widely used tools in computer algebra, a claim which is 
supported, for instance, by the 3400+ publications currently listed in the online \emph{Gr\"obner 
Bases Bibliography}.\footnote{\url{http://www.risc.jku.at/Groebner-Bases-Bibliography/}} Their 
importance stems from the fact that they generalize, at the same time, Gauss' algorithm for solving 
systems of linear equations and Euclid's algorithm for computing the GCD of univariate polynomials: 
Gr\"obner bases enable the effective, systematic solution of a variety of problems in polynomial 
ideal theory, ranging from the decision of ideal membership and ideal congruence, the solution 
of systems of algebraic equations, to as far as automatic theorem proving. Since it is clearly beyond the 
scope of this paper to mention all the merits, applications, and generalizations of Gr\"obner bases, 
we refer the interested reader to any standard textbook about Gr\"obner bases (e.\,g. 
\cite{Adams94,Kreuzer00}) instead; nonetheless, Section~\ref{sec::Background} briefly presents the 
mathematical background of Gr\"obner bases.

The main achievement we report on in this paper is the first-time formalization of said theory in 
the proof assistant Isabelle/HOL~\cite{Nipkow2002}. Although Gr\"obner bases have been formalized 
in other proof assistants already (a list of which can be found in Section~\ref{sec::Related}), our 
work features the, to the best of our knowledge, first computer-certified implementation of 
Faug\`ere's $F_4$ algorithm~\cite{Faugere99} for computing Gr\"obner bases by matrix reductions, as 
well as the (again to the best of our knowledge) first-time formal treatment of the theory in the 
more general setting of \emph{modules} and \emph{submodules} rather than rings and 
ideals~\cite{Kreuzer00}.

Summarizing, the highlights of our elaboration are
\begin{itemize}
 \item an abstract view of power-products that does not refer to the notion of ``indeterminate'' at 
all and that allows us to represent power-products by functions of type \type{nat\,$\Rightarrow$\,nat} 
with finite support (Section~\ref{sec::PowerProducts}),

 \item an abstract view of vectors of polynomials, that can be interpreted both by ordinary scalar polynomials and by functions mapping indices (of the components) to scalar polynomials (Section~\ref{sec::Polynomials}),
 
 \item the definition of Gr\"obner bases via confluence of the reduction relation they induce (definition~\thm{is-Groebner-basis} in Section~\ref{sec::GB}),
 
 \item the proof of the main theorem about S-polynomials (theorem~\thm{Buchberger-thm-finite} in Section~\ref{sec::GB}),
 
 \item an alternative characterization of Gr\"obner bases via divisibility of leading 
terms (theorem~\thm{GB-alt-finite} in Section~\ref{sec::GB}),

 \item a generic algorithm schema for computing Gr\"obner bases, of which both Buchberger's algorithm and Faug\`ere's $F_4$ algorithm are instances (function \const{gb\us schema} in Section~\ref{sec::Schema}),
 
 \item the implementation of Buchberger's criteria for increasing the efficiency of said algorithm schema (Section~\ref{sec::Criteria}),
 
 \item the formally verified implementation of the $F_4$ algorithm (Section~\ref{sec::F4}),
 
 \item the definition and a constructive proof of existence and uniqueness of reduced Gr\"obner 
bases (Section~\ref{sec::Reduced}), and

 \item a formally verified algorithm for computing Gr\"obner bases of syzygy modules (Section~\ref{sec::Syzygy}),
 
 \item the proper set-up of Isabelle's code generator to produce certified executable code (Section~\ref{sec::Code}).
\end{itemize}

An Isabelle2017-compatible version of the formalization presented in this paper is available 
online~\cite{Formalization}. Furthermore, a big portion of the formalization has already been added 
to the development version of the Archive of Formal Proofs (AFP). Note also that there is a 
Gr\"obner-bases entry in the release version of the AFP~\cite{Groebner-Bases-AFP} (which will be 
replaced by the one in the development version upon the next release of Isabelle), but it lacks many 
features compared to~\cite{Formalization}.


\subsection{Related Work}
\label{sec::Related}

Gr\"obner bases have been formalized in a couple of other proof assistants already. The first formalizations date back to around 2000, when Th\'ery~\cite{Thery2001} and Persson~\cite{Persson2001} formalized basically the same aspects in Coq~\cite{Bertot2004} that we recently formalized in Isabelle/HOL (except $F_4$ and modules). The presentation of their theory is on a fairly abstract level, similar to our case.
Moreover, it is possible to automatically extract executable, certified OCaml code for computing Gr\"obner bases from the formalization. In 2009, Jorge, Guilas and Freire~\cite{Jorge2009} took the reverse direction: they \emph{first} implemented an efficient version of Buchberger's algorithm directly in OCaml and \emph{then} proved it correct, making use of the underlying formal 
theory in Coq.

Another formalization that focuses very much on the actual \emph{computation} of Gr\"obner bases is that of Medina-Bulo, Palomo-Lozano, Alonso-Jim\'enez and Ruiz-Reina~\cite{Medina-Bulo2010} in ACL2~\cite{Kaufmann2000}, dating back to 2010. There, however, the representation of power-products and polynomials is fixed to ordered lists of exponents and monomials, respectively, owing to the limited expressiveness of the underlying system.

In 2006, Schwarzweller~\cite{Schwarzweller2006} formalized Gr\"obner bases in Mizar~\cite{Bancerek2015}. He also dealt with polynomial reduction, Buchberger's algorithm and reduced Gr\"obner bases, and in addition proved some other equivalent characterizations of Gr\"obner bases (e.\,g. via so-called \emph{standard representations} of polynomials).

Very recently, the first author formalized a generalization of Gr\"obner bases~\cite{Maletzky2016} in the proof assistant Theorema~2.0~\cite{Buchberger2016}. His work is not confined to polynomial rings over fields, but considers much wider classes of commutative rings where Gr\"obner bases can be defined and computed (so-called \emph{reduction rings}). All certified algorithms are directly executable within Theorema.
In the same system, Buchberger~\cite{Buchberger2004} and Craciun~\cite{Craciun2008} took Buchberger's algorithm as a case study for the automatic synthesis of algorithms: they managed to synthesize the algorithm only from its specification by the so-called \emph{lazy thinking} method.

Apart from the formalizations listed above, Gr\"obner bases have been successfully employed by various proof assistants (among them HOL~\cite{Harrison2001} and Isabelle/HOL~\cite{Chaieb2007}) as proof methods for proving universal propositions over rings. In a nutshell, this proceeds by showing that the system of polynomial equalities and inequalities arising from refuting the original formula is unsolvable, which in turn is accomplished by finding a combination of these polynomials that yields a non-zero constant polynomial -- and this is exactly what Gr\"obner bases can do. The computation of Gr\"obner bases, however, is taken care of by a ``black-box'' \textsc{ml} program whose correct behavior is irrelevant for the correctness of the overall proof step, since the obtained witness is independently checked by the trusted inference kernel of the system. The work described in this paper is orthogonal to~\cite{Chaieb2007} in the sense that it formalizes the theory underlying Gr\"obner bases and proves the 
total correctness of the algorithm, which is not needed in~\cite{Chaieb2007}.

Our work builds upon existing formal developments of multivariate polynomials~\cite{Polynomials-AFP} (to which we also contributed) and abstract rewrite systems~\cite{Abstract-Rewriting-AFP}, and $F_4$ in addition builds upon Gauss-Jordan normal forms of matrices~\cite{Jordan_Normal_Form-AFP}.

\section{Mathematical Background}
\label{sec::Background}

We now give a very brief overview of the mathematical theory of Gr\"obner bases for (sub)modules, 
only to keep this paper self-contained. The interested reader is referred to any standard textbook 
about Gr\"obner bases, e.\,g. \cite{Adams94,Kreuzer00}, for a more thorough account on the 
subject; in particular, \cite{Kreuzer00} also presents Gr\"obner bases for modules. Readers 
familiar with Gr\"obner bases in rings but not with Gr\"obner bases in modules will spot only few 
differences to the ring-setting, as the module-setting parallels the other in many respects.

The theory of Gr\"obner bases for modules is concerned with vectors of commutative multivariate 
polynomials over fields, and more precisely with effectively solving \emph{module-theoretic} 
problems. Hence, let in the remainder of this section $K$ be some field, $X=\{x_0,\ldots,x_{n-1}\}$ 
be a finite set of indeterminates, $k\neq 0$ be a natural number, and let 
$K[X]^k=K[x_0,\ldots,x_{n-1}]^k$ denote the $k$-dimensional free module over the (commutative) ring 
of $n$-variate polynomials over $K$. We will refer to products of indeterminates as 
\emph{power-products} (e.\,g. $x_0$, $x_2^5x_3^2$); the set of all power-products in the 
indeterminates $X$ is denoted by $[X]$. The $k$ elements of the canonical basis of $K^k$ are
denoted by $e_j$, for $0\leq j<k$; hence, every element $p$ of $K[X]^k$ can 
be uniquely written as a sum $p=\sum_{j=0}^k p_j\,e_j$ for polynomials $p_j\in K[X]$. \emph{Terms} 
are polynomials of the form $t\,e_j$ for a power-product $t$ and $0\leq j<k$; their importance stems 
from the fact that the set of terms, denoted by $[X]^k$, is a basis of the infinite-dimensional 
$K$-vector space $K[X]^k$.

\begin{example}
\label{exm::vecpoly0}
Let $p=\left(\begin{array}{c}x_1^2-x_0x_1\\2x_0+3\end{array}\right)\in\mathbb{Q}[x_0,x_1]^2$. Then 
$p$ can be written as a linear combination of terms as
\[
p=
1\cdot x_1^2\,\underbrace{\left(\begin{array}{c}1\\0\end{array}\right)}_{=e_0}+
(-1)\cdot x_0x_1\,\left(\begin{array}{c}1\\0\end{array}\right)+
2\cdot x_0\,\underbrace{\left(\begin{array}{c}0\\1\end{array}\right)}_{=e_1}+
3\cdot 1\,\left(\begin{array}{c}0\\1\end{array}\right).
\]
\end{example}

\subsection{Gr\"obner Bases}

First of all, we now have to choose an \emph{admissible order} on the power-products: a linear 
order $\preceq$ on power-products is called \emph{admissible} iff $1$ is the smallest element and 
if $s\preceq t$ implies $s\cdot u\preceq t\cdot u$, for all $s,t,u\in [X]$. An example of such an 
order is the purely lexicographic order, which compares two power-products $s,t$ by successively 
comparing the exponents of the $x_i$ in $s$ and $t$ until some are non-equal. In the remainder of 
this section we fix an admissible order $\preceq$.

In the next step, we must extend $\preceq$ to an ordering $\ordv$ on terms. As before, we have some 
freedom in doing so, as long as (i)~$s\,e_i\ordv t\,e_j$ implies $(u\cdot s)\,e_i\ordv (u\cdot 
t)\,e_j$ for all $s,t,u\in[X]$, and (ii)~$s\preceq t\wedge i\leq j$ implies $s\,e_i\ordv t\,e_j$ for 
all $s,t\in[X]$. Examples of $\ordv$ are \emph{position-over-term} (POT) and 
\emph{term-over-position} orders: in a POT order $\ordv^{\sf pot}$ we have $s\,e_i\ordv^{\sf pot} 
t\,e_j$ iff $i<j\vee (i=j\wedge s\preceq t)$, whereas in a TOP order $\ordv^{\sf top}$ we have 
$s\,e_i\ordv^{\sf top} t\,e_j$ iff $s\prec t\vee (s=t\wedge i\leq j)$. Hence, both POT and TOP are 
lexicographic combinations of $\preceq$ and $\leq$, and easily seen to satisfy the two requirements 
listed above.

Having $\ordv$, we can define the notions of \emph{leading term}, \emph{leading power-product} and 
\emph{leading coefficient} of non-zero vector-polynomials $p\in K[X]^k$: the leading term of $p$, 
written $\lt(p)$, is simply the largest term w.\,r.\,t. to $\ordv$ that appears in $p$ with 
non-zero coefficient, the leading power-product $\lp(p)$ is the power-product of the leading term, 
and the leading coefficient $\lc(p)$ is the coefficient of $\lt(p)$ in $p$. In general, the 
coefficient of a term $v$ in a polynomial $p$ is denoted by $\C(p,v)$.
\begin{example}
Let $p$ as in Example~\ref{exm::vecpoly0}, and assume $\preceq$ is the lexicographic order relation 
with $x_0\prec x_1$. With $\ordv^{\sf pot}$ we obtain $\lt(p)=x_0\,e_1$ and $\lc(p)=2$; with 
$\ordv^{\sf top}$, $\lt(p)=x_1^2\,e_0$ and $\lc(p)=1$.
\end{example}

\begin{definition}[Reduction]
\label{def::Reduction}
Let $p,q,f\in K[X]^k$ with $f\neq 0$ and $t\in [X]$. Then $p$ \emph{reduces to} $q$ \emph{modulo} 
$f$ \emph{using} $t$, written $p\rightarrow_{f,t} q$, iff $\lt(f)=s\,e_j$, $\C(p,(t\cdot 
s)\,e_j)\neq 0$ and $q=p-\frac{\C(p,(t\cdot s)\,e_j)}{\lc(f)}\cdot t\cdot f$.\\
If $F\subseteq K[X]^k$ then we write $p\rightarrow_F q$ iff there exist $f\in F\backslash\{0\}$ and 
$t\in [X]$ with $p\rightarrow_{f,t} q$. As usual, $\rightarrow_F^*$ denotes the 
reflexive-transitive closure of $\rightarrow_F$.
\end{definition}

For any $F\subseteq K[X]$, $\rightarrow_F$ can be shown to be terminating, i.\,e. there do not 
exist infinite chains of reductions. However, in general $\rightarrow_F$ is not \emph{confluent}, as 
can be seen in the following example:

\begin{example}[Example~2.5.7. in~\cite{Kreuzer00}]
\label{ex::not-confluent}
Let $F=\{f_1,f_2\}\subseteq \mathbb{Q}[x_0,x_1]\cong \mathbb{Q}[x_0,x_1]^1$ where $f_1=x_1^2$ and 
$f_2=x_0x_1+x_0^2$, and assume $\preceq$ and $\ordv$ are the lexicographic order with $x_0\prec 
x_1$. Then $x_0x_1^2\rightarrow_{f_1,x_0}0$ and 
$x_0x_1^2\rightarrow_{f_2,x_1}-x_0^2x_1\rightarrow_{f_2,x_0}x_0^3$; both $0$ and $x_0^3$ are 
irreducible modulo $F$, so $\rightarrow_F$ is not confluent.
\end{example}

The observation that $\rightarrow_F$ is in general not confluent motivates the following

\begin{definition}[Gr\"obner basis]
\label{def::GB}
A set $G\subseteq K[X]^k$ is a \emph{Gr\"obner basis} iff $\rightarrow_G$ is confluent.
\end{definition}

Note that the notion of reduction, and hence also that of Gr\"obner basis, strongly depends on the 
implicitly fixed term order $\ordv$!

If a set $F$ is no Gr\"obner basis, it can be \emph{completed} to one by successively considering 
all \emph{critical pairs} and adding new elements to the set that ensure that all critical pairs 
have a common successor w.\,r.\,t. the reduction relation, just as in the well-known Knuth-Bendix 
procedure for general term rewrite systems. In the case of multivariate polynomials the algorithm is 
called \emph{Buchberger's algorithm} and has the nice property that it always terminates for any 
finite input $F$.

\begin{definition}[S-polynomial]
\label{def::SPoly}
Let $f,g\in K[X]^k\backslash\{0\}$, and assume $\lt(f)=s\,e_i$ and $\lt(g)=t\,e_j$. Then the 
\emph{S-polynomial} of $f$ and $g$, written 
$\spoly(f,g)$, is defined as
\[
\spoly(f,g):=\left\{\begin{array}{c c c}
\frac{\const{lcm}(s,t)}{\lc(f)\cdot s}\cdot f-
\frac{\const{lcm}(s,t)}{\lc(g)\cdot t}\cdot g & \Leftarrow & i=j\\
0 & \Leftarrow & i\neq j
\end{array}\right..
\]
\end{definition}

The S-polynomial of $f$ and $g$ is precisely the difference of the critical pair of $f$ and $g$, so 
it roughly corresponds to the smallest element where reduction modulo $\{f,g\}$ might diverge. In 
the usual Knuth-Bendix procedure one reduces the two constituents of a critical pair individually 
and then checks whether the normal forms are equal; in our case, it suffices to first compute their 
difference (i.\,e. the S-polynomial), then reduce the S-polynomial to normal form, and finally check 
whether the normal form is $0$. This idea is summarized in the following

\begin{theorem}[Buchberger, 1965]
\label{thm::SPoly}
Let $G\subseteq K[X]^k$. Then $G$ is a Gr\"obner basis iff for all $f,g\in G$, 
$\spoly(f,g)\rightarrow_G^* 0$.
\end{theorem}

Therefore, completing a finite set $F$ to a Gr\"obner basis simply proceeds by
\begin{enumerate}
 \item forming all S-polynomials of elements in the current set,
 \item reducing them to normal form modulo the current set,
 \item adding those normal forms that are not $0$ to the current set (which means that they can be 
further reduced to $0$ modulo the enlarged set, since we always have $p\rightarrow_{p,1} 0$), and
 \item repeating this procedure until all S-polynomials can be reduced to $0$.
\end{enumerate}

This procedure is called \emph{Buchberger's algorithm}, which is justified by

\begin{theorem}[Buchberger, 1965]
\label{thm::Term}
The procedure outlined above terminates after finitely many iterations, for any finite input set 
$F$ and admissible term order $\ordv$, and regardless in which order and how the S-polynomials 
are 
reduced.
\end{theorem}

\begin{example}
\label{ex::GB}
Let $F$, $f_1$, $f_2$, $\preceq$ and $\ordv$ be as in Example~\ref{ex::not-confluent}. We 
apply Buchberger's algorithm to compute a Gr\"obner basis of $F$.\\
We start with the S-polynomial of $f_1$ and $f_2$:
\[
\spoly(f_1,f_2)=x_0\,f_1-x_1\,f_2=-x_0^2x_1\rightarrow_{f_2,x_0}x_0^3.
\]
$f_3=x_0^3$ is irreducible modulo $F$, so we must add it to $F$: $\tilde{F}=F\cup\{f_3\}$. This 
ensures that $\spoly(f_1,f_2)$ can be reduced to $0$ modulo the enlarged set $\tilde{F}$, but it 
also means that we have to consider $\spoly(f_1,f_3)$ and $\spoly(f_2,f_3)$ as well.
\[
\spoly(f_1,f_3)=x_0^3\,f_1-x_1^2\,f_3=0.
\]
Since the S-polynomial of $f_1$ and $f_3$ is $0$, we do not have to augment $\tilde{F}$ by a new 
polynomial.
\[
\spoly(f_2,f_3)=x_0^2\,f_2-x_1\,f_3=x_0^4\rightarrow_{f_3,x_0}0.
\]
Since the S-polynomial of $f_2$ and $f_3$ can be reduced to $0$ modulo $\tilde{F}$, and 
therefore \emph{all} S-polynomials can be reduced to $0$, $\tilde{F}$ is a Gr\"obner basis of $F$.
\end{example}

\subsection{Submodules}

We assume familiarity with the concept of a \emph{submodule} of a module $M$, as a subset of $M$ 
that is closed under addition and under multiplication by arbitrary elements from $M$. Regarding 
notation, we write $\const{pmdl}(F)\subseteq M$ for the submodule generated by the set $F\subseteq 
M$. In our case, $M$ is of course $K[X]^k$; if $k=1$, submodules are nothing else than ideals.

As can be easily seen, Buchberger's algorithm preserves the submodule generated by the set in 
question, i.\,e. if Buchberger's algorithm applied to $F$ yields $G$, then $\const{pmdl}(F)=\const{pmdl}(G)$.
Hence, we can conclude that every finitely generated submodule of 
$K[X]^k$ has a finite Gr\"obner basis; this Gr\"obner basis is not unique, though: first of all it 
clearly depends on $\ordv$, but even if $\ordv$ is fixed, $G$ is a Gr\"obner basis (w.\,r.\,t. 
$\ordv$) and $H\subseteq\const{pmdl}(G)$, then $G\cup H$ is still a Gr\"obner basis of $\const{pmdl}(G)$.

One important property of Gr\"obner bases $G$ is that the unique normal form w.\,r.\,t. 
$\rightarrow_G$ of an arbitrary polynomial $p$ is $0$ iff $p\in\const{pmdl}(G)$. So, since normal 
forms are effectively computable, also the \emph{submodule membership problem} is effectively 
decidable if the submodule $N$ in question is given by a finite generating set $F$: just compute a 
Gr\"obner basis $G$ of $F$ by Buchberger's algorithm, reduce the polynomial in question to normal 
form w.\,r.\,t. $G$, and check whether the result is $0$.

\begin{example}
Continuing Example~\ref{ex::GB}: $x_0^5\in\const{pmdl}(F)$, because $x_0^5\rightarrow_{f_3,x_0^2} 
0$ and $f_3$ is contained in the Gr\"obner basis $\tilde{F}$ computed from $F$. However, $x_0^5$ is 
irreducible modulo $F$ itself, illustrating that it is really crucial to perform the reduction 
modulo a \emph{Gr\"obner basis} when deciding submodule membership, not just modulo \emph{any} 
generating set.
\end{example}

\section{Multivariate Polynomials}
\label{sec::MVPolynomials}

Since Gr\"obner bases are concerned with (vectors of) multivariate polynomials, we have to spend 
some words on the formalization of such polynomials in Isabelle/HOL.

The formal basis of multivariate polynomials are so-called \emph{polynomial mappings}, originally 
formalized by Haftmann \emph{et~al.}\ \cite{Haftmann2014}, extended by 
Bentkamp~\cite{Deep-Learning-AFP}, and now part of the AFP-entry 
\thy{Polynomials}~\cite{Polynomials-AFP} in the development version of the Archive of Formal 
Proofs. A polynomial mapping is simply a function of type $\alpha\Rightarrow\beta\class{::zero}$ 
with finite support, i.\,e., all but finitely many arguments are mapped to 
$0$.\footnote{$\beta\class{::zero}$ is a type-class constraint on type $\beta$, stipulating that 
there must be a distinguished constant $0$ of type $\beta$. See~\cite{TypeClass} for information on 
type classes in Isabelle/HOL.} In Isabelle/HOL, as well as in the remainder of this paper, the type 
of polynomial mappings is called \type{poly\us mapping} and written in infix form as 
$\polyab$, where $\beta$ is tacitly assumed to belong to type class \class{zero}. Formally, 
\type{poly\us mapping} is defined as
\begin{lstlisting}
typedef (overloaded) ($\alpha$, $\beta$) poly%\us%mapping = "{f::$\alpha\Rightarrow\beta$::zero. finite {x. f x $\neq$ 0}}"
\end{lstlisting}

The importance of type \type{poly\us mapping} stems from the fact that not only polynomials, but 
also power-products (i.\,e. products of indeterminates, like $x_0^3 x_1^2$) can best be thought of 
as terms of this type: in power-products, indeterminates are mapped to their exponents (with only 
finitely many being non-zero), and in polynomials, power-products are mapped to their coefficients 
(again only finitely many being non-zero). Hence, a scalar polynomial would typically be a term of 
type $\polym{(\polym{\chi}{\type{nat})}}{\beta}$, with $\chi$ being the type of the 
indeterminates and $\beta$ being the type of the coefficients.

\begin{remark}
\label{rem::TypeClass}
Although the mathematical theory our formalization is concerned with clearly 
belongs to the field 
of Algebra, we did not follow HOL-Algebra's approach to algebraic structures 
with explicit carrier 
sets, but instead based everything on \emph{type classes}. In particular, the 
coefficient type of 
polynomials must belong to type class \class{field}, and the type of 
power-products must belong to 
the custom-made type class \class{graded\us dickson\us powerprod} -- this 
already indicates that we 
developed the theory more abstractly than fixing the type of power-products to 
something like 
$\polym{\chi}{\type{nat}}$; see below for 
details.
\end{remark}

\subsection{Power-Products}
\label{sec::PowerProducts}

Instead of fixing the type of power-products to $\polym{\chi}{\type{nat}}$ throughout the 
formalization, we opted to develop the theory slightly more abstractly: power-products can be of 
\emph{arbitrary} type, as long as the type belongs to a certain type class that allows us to prove 
all key results of the theory. Said type class is called \class{graded\us dickson\us powerprod} and 
defined as
\begin{lstlisting}
class graded%\us%dickson%\us%powerprod = cancel%\us%comm%\us%monoid%\us%mult + dvd +
  fixes lcm :: "$\alpha\Rightarrow\alpha\Rightarrow\alpha$"
  assumes dvd%\us%lcm: "s dvd (lcm s t)"
  assumes lcm%\us%dvd: "s dvd u $\Longrightarrow$ t dvd u $\Longrightarrow$ (lcm s t) dvd u"
  assumes lcm%\us%comm: "lcm s t = lcm t s"
  assumes times%\us%eq%\us%one: "s * t = 1 $\Longrightarrow$ s = 1"
  assumes ex%\us%dgrad: "$\exists$d::$\alpha\Rightarrow$%\,%nat. dickson%\us%grading d"
\end{lstlisting}

Several remarks on the above definition are in order. First of all, note that the base class of 
\class{graded\us dickson\us powerprod} is the class of \emph{cancellative commutative multiplicative 
monoids} which in addition feature a \emph{divisibility relation} (infix \const{dvd}). Furthermore, 
types belonging the the class must also
\begin{itemize}
 \item provide a function called \const{lcm} that possesses the usual properties of \emph{least 
common multiple},
 \item obey the law that a product of two factors can only be $1$ if both factors are $1$, i.\,e. 
$1$ is the only invertible element, and
 \item admit a so-called \emph{Dickson grading}.
\end{itemize}
Dickson gradings are a technicality we need for being able to represent power-products conveniently 
as functions of type $\polym{\type{nat}}{\type{nat}}$ (with an infinite supply of indeterminates) in 
actual computations, without having to having to care how many indeterminates 
actually appear in the computations.\footnote{This representation is also suggested 
in~\cite{Haftmann2014}.} Gr\"obner bases normally only work with a fixed finite set of 
indeterminates, for otherwise the reduction relation $\rightarrow_F$ of 
Definition~\ref{def::Reduction} is not terminating in general. Therefore, if we want to work with 
potentially infinitely many indeterminates, we need a means to ensure that only finitely many 
appear with non-zero exponent and non-zero coefficient in possibly infinite sets of polynomials -- 
and this is exactly what a Dickson grading $d$ does. In fact, $d$ can best be thought of giving, 
for an abstract power-product $t$, the ``index'' of the highest indeterminate occurring in $t$ with 
non-zero exponent. So, for ensuring that only finitely many indeterminates appear in a set $F$ of 
polynomials, it suffices to stipulate the existence of some natural number $m$ such that $d(t)\leq 
m$ for all power-products $t$ appearing in $F$. Dickson gradings are called such because they 
ensure, by definition of \const{dickson\us grading}, the \emph{Dickson property}~\cite{Dickson13} 
of sequences of abstract power-products:
\begin{lstlisting}
lemma dickson%\us%property:
  fixes s::"nat $\Rightarrow$ $\alpha$"
  assumes "dickson%\us%grading d" and "$\bigwedge$i. d (s i) $\leq$ d (s 0)"
  obtains i j where "i < j" and "(s i) dvd (s j)"
\end{lstlisting}
Thus, in any sequence $s$ of abstract power-products in which only finitely many indeterminates 
appear (second assumption of \thm{dickson-property}), there exist indices $i<j$ such that $s_i$ 
divides $s_j$. In other words, the divisibility relation is some sort of \emph{well-quasi order} on 
power-products. Dickson's lemma, finally, states that such Dickson 
gradings indeed exist for type $\polym{\type{nat}}{\type{nat}}$: in a 
power-product of type $\polym{\type{nat}}{\type{nat}}$, just take the 
largest number that is not mapped to $0$.\footnote{Well-quasi orders and Dickson's lemma have been 
formalized in Isabelle/HOL already~\cite{Well-Quasi-Orders-AFP}, but we proved the lemma in our 
(slightly more general) setting from scratch.}

\begin{remark}
In the actual formalization, power-products are written additively rather than multiplicatively: 
the base class is \class{cancel\us comm\us monoid\us add}, the monoid operation is $+$, the neutral 
element is $0$, the least common multiple is called \const{lcs} (standing for ``least common sum'') 
and the divisibility relation is called \const{adds}. The reason for doing so is a mere 
technicality: it better integrates with the existing 
type-class hierarchy of Isabelle/HOL, allowing us to reuse existing point-wise instantiations of the 
various group- and ring-related type classes by the function type $\Rightarrow$.
\end{remark}

Finally, in order to formalize Gr\"obner bases we need to fix an 
admissible order relation on power-products, as explained in Section~\ref{sec::Background}. 
Since there are infinitely 
many of them we do not want to restrict the formalization to a particular one 
but instead parametrized all subsequent definitions, theorems and algorithms 
over such a relation through the use of a \emph{locale}~\cite{Ballarin2010} (note that we have to 
provide both the reflexive and the strict version of the relation):
\begin{lstlisting}
locale gd%\us%powerprod =
  linorder ord ord%\us%strict
  for ord::"$\alpha\Rightarrow\alpha$::graded%\us%dickson%\us%powerprod $\Rightarrow$ bool" (infix "$\preceq$" 50)
  and ord_strict (infixl "$\prec$" 50) +
  assumes one%\us%min: "1 $\preceq$ t"
  assumes times%\us%monotone: "s $\preceq$ t $\Longrightarrow$ s * u $\preceq$ t * u"
\end{lstlisting}

In addition, the formalization features three concrete orders on type 
$\polym{\type{nat}}{\type{nat}}$ that are proved to be admissible: the purely lexicographic order 
\const{lex}, the degree-lexicographic order \const{dlex}, and the degree-reverse-lexicographic 
order \const{drlex}. Each of these orders can without any further ado be used in certified 
computations of Gr\"obner bases in Isabelle/HOL.

\subsection{Polynomials}
\label{sec::Polynomials}

Having described our formalization of power-products, we now turn to 
polynomials. In fact, most definitions related to, and facts about, multivariate polynomials as 
objects of type $\polyab$ that are required by our Gr\"obner 
bases formalization were already formalized~\cite{Haftmann2014,Deep-Learning-AFP}: addition, 
multiplication, coefficient-lookup 
(called \const{lookup} in the formal theories but denoted by the more intuitive \C\ in the 
remainder) and support (called \const{keys}). These things are all pretty much standard, so we do 
not go into more detail here. We only emphasize that henceforth $\alpha$ is the type of 
power-products, i.\,e. is tacitly assumed to belong to type-class \class{graded\us dickson\us 
powerprod}.

What is certainly more interesting is the way how we represent vectors of polynomials: since we 
formulate the theory of Gr\"obner bases for free modules over polynomial rings over fields, i.\,e. 
for structures of the form $K[x_0,\ldots,x_n]^k$, we need to specify what the formal type of such 
structures is in our formalization. Going back to Section~\ref{sec::Background} one realizes that 
the best way (for our purpose) to represent vector-polynomials is as $K$-linear combinations of 
terms $t\,e_i$. Or, in other words, as polynomial mappings mapping terms to coefficients. Terms, in 
turn, can most conveniently be represented as pairs of power-products (like $t$) and 
component-indices (like $i$), without having to care what exactly the basis elements $e_i$ are. 
Therefore, the formal type of vector polynomials is $\polyakb$, where $\alpha$ and $\beta$ have 
their usual meaning as type of power-products and coefficients, respectively, and $\kappa$ is the 
type of component-indices.
\begin{example}
\label{exm::vecpoly}
Let $p$ be as in Example~\ref{exm::vecpoly0}. Then $p$ is represented by the polynomial mapping 
which maps the pair $(x_1^2,0)$ to $1$, $(x_0x_1,0)$ to $-1$, $(x_0,1)$ to $2$, $(1,1)$ to $3$, and 
all other pairs to $0$.
\end{example}
As Example~\ref{exm::vecpoly} shows, for representing a $k$-dimensional vector $\kappa$ does not 
need to have exactly $k$ elements, but only \emph{at least} $k$ elements. This saves us from 
introducing dedicated types for $1$, $2$, $3$, \ldots\ dimensions, as we can use \type{nat} 
throughout. Nonetheless, we do not fix $\kappa$ to \type{nat}, because in some situations it is 
desirable to restrict definitions or theorems to one-dimensional (scalar) polynomials, while 
still building upon concepts defined for vector-polynomials and arbitrary $\kappa$. This can be 
achieved easily by instantiating $\kappa$ by the unit type \type{unit}.\footnote{\type{unit} 
contains only one single element, so $\alpha\times\type{unit}$ is isomorphic to $\alpha$.}

\begin{remark}
If $\kappa$ is instantiated by \type{nat} we get similar problems as with infinitely many 
indeterminates and, hence, must provide similar means for ensuring that only finitely many 
components in (infinite) sets of infinite-dimensional polynomials are non-zero. We omit the 
(not so intricate) details here.
\end{remark}

Let us now turn to the extension $\ordv$ of $\preceq$: since $\preceq$ can be extended in many 
different ways to $\ordv$, and we do not want to restrict ourselves to a particular choice, we 
again employ a locale for parametrizing all subsequent definitions and lemmas over any admissible 
instance of $\ordv$:
\begin{lstlisting}
locale gd%\us%term =
  gd%\us%powerprod ord ord%\us%strict +
  ord%\us%term%\us%lin: linorder ord%\us%term ord%\us%term%\us%strict
    for ord :: "$\alpha\Rightarrow\alpha::\class{graded\us dickson\us powerprod}\Rightarrow$ bool" (infix "$\preceq$" 50)
    and ord%\us%strict (infix "$\prec$" 50)
    and ord%\us%term :: "$(\alpha\times\kappa)\Rightarrow(\alpha\times\kappa::\class{wellorder})\Rightarrow$%\,%bool" (infix "$\ordv$" 50)
    and ord%\us%term%\us%strict (infix "$\ordsv$" 50) +
  assumes stimes%\us%mono: "v $\ordv$ w $\Longrightarrow$ t $\circledast$ v $\ordv$ t $\circledast$ w"
  assumes ord%\us%termI: "fst v $\preceq$ fst w $\Longrightarrow$ snd v $\leq$ snd w $\Longrightarrow$ v $\ordv$ w"
\end{lstlisting}
So, \locale{gd\us term} extends \locale{gd\us powerprod} by $\ordv$ and $\ordsv$, requires $\kappa$ to be well-ordered by $\leq$, and requires $\ordv$ to be a linear ordering satisfying the two axioms \thm{stimes-mono} and \thm{ord-termI}. $\circledast$ is defined as
\begin{lstlisting}
definition stimes :: "$\alpha\Rightarrow(\alpha\times\kappa)\Rightarrow(\alpha\times\kappa)$" (infixl "$\circledast$" 75)
  where "stimes t v = (t * fst v, snd v)"
\end{lstlisting}
i.\,e. it multiplies the power-product of its second argument with its first argument. \const{fst} 
and \const{snd} are built-in Isabelle/HOL functions that access the first and second, respectively, 
component of a pair. In the formalization, we prove two interpretations of the locale: one for POT 
orders, and one for TOP orders.

In the context of the locale, i.\,e. with $\ordv$ as an implicit parameter available, we can now 
immediately define leading terms, leading power-products and leading coefficients, just as 
described in Section~\ref{sec::Background}. In addition, we also define the \emph{tail} of a 
polynomial $p$, $\const{tail}(p)$, as a copy of $p$ where, however, the leading coefficient of $p$ 
is set to $0$, i.\,e. in which only terms less than $\lt(p)$ appear.

Based on $\circledast$ we introduce multiplication of a vector-polynomial by a coefficient 
$c::\beta$ and a power-product $t::\alpha$ in the obvious way: all coefficients are multiplied by 
$c$, and all terms are multiplied by $t$ via $\circledast$. The resulting function is called 
\const{monom\us mult} and caters for the multiplications needed in reduction (cf. 
Definition~\ref{def::Reduction}) and S-polynomial (cf. Definition~\ref{def::SPoly}).

Before we finish this section, we introduce three more notions related to vector-polynomials that 
we are going to need later. First, we extend divisibility of power-products of type $\alpha$ to 
divisibility of terms of type $\alpha\times\kappa$:
\begin{lstlisting}
definition dvd%\us%term :: "$(\alpha\times\kappa)\Rightarrow(\alpha\times\kappa)\Rightarrow$ bool" (infix "$\dvdv$" 50)
  where "dvd%\us%term u v $\longleftrightarrow$ (snd u = snd v $\wedge$ (fst u) dvd (fst v))"
\end{lstlisting}
which means that a term $s\,e_i$ divides another term $t\,e_j$ iff $i=j$ and $s$ divides $t$.

Furthermore, we extend the linear order $\ordv$ to a partial order $\ordp$ on polynomials, whose 
strict version $\ordsp$ is defined as
\begin{lstlisting}
definition ord%\us%strict%\us%p::"$(\polyakb)\Rightarrow(\polyakb)\Rightarrow$%\,%bool" (infix "$\ordsp$" 50)
  where "p $\ordsp$ q $\longleftrightarrow$ ($\exists$v. coeff p v = 0 $\wedge$ coeff q v $\neq$ 0 $\wedge$
                          ($\forall$u. v $\ordsv$ u $\longrightarrow$ coeff p u = coeff q u))"
\end{lstlisting}
and prove that $\ordsp$ is well-founded on sets of polynomials in which only finitely many 
indeterminates appear:
\begin{lstlisting}
lemma ord%\us%p%\us%minimum:
  assumes "dickson%\us%grading d" and "x $\in$ Q" and "Q $\subseteq$ dgrad%\us%p%\us%set d m"
  obtains q where "q $\in$ Q" and "$\bigwedge$y. y $\ordsp$ q $\Longrightarrow$ y $\notin$ Q"
\end{lstlisting}
$\const{dgrad\us p\us set}(d,m)$ gives the set $F$ of all polynomials such that the Dickson grading 
$d$ only attains values below $m$ when applied to power-products appearing in $F$. Hence, informally
$Q\subseteq\const{dgrad\us p\us set}(d,m)$ expresses that at most $m$ indeterminates 
appear in $Q$.

Finally, we also introduce the notion of a \emph{submodule} generated by a set $B$ of polynomials 
as the smallest set that contains $B\cup\{0\}$ and that is closed under addition and under 
\const{monom\us mult}:
\begin{lstlisting}
inductive_set pmdl :: "$\polyakb$ set $\Rightarrow$ $\polyakb$ set" for B where
  pmdl%\us%0: "0 $\in$ pmdl B"|
  pmdl%\us%plus: "a $\in$ pmdl B $\Longrightarrow$ b $\in$ B $\Longrightarrow$ a + monom%\us%mult c t b $\in$ pmdl B"
\end{lstlisting}
Submodules generalize the concept of \emph{ideals} in rings, which are sets that are closed under 
addition and under multiplication by arbitrary elements of the ring.

\begin{remark}
Although in the theory of Gr\"obner bases polynomials need to have coefficients in fields, we 
formulated all definitions and lemmas about polynomials for as general coefficient 
types as possible, often requiring only the ubiquitous type-class constraint \class{zero} (which we omit 
in this paper for better readability).
\end{remark}

\begin{remark}
Developing the theory for modules and submodules is a nice generalization of rings and ideals, and 
as mentioned before it is always possible to instantiate $\kappa$ by \type{unit} when one wishes to 
state definitions/theorems for scalar polynomials only. But still, all types would then involve the 
redundant artifact \type{unit}, which would clutter the theory quite a bit. As a remedy, we actually 
do not fix the type $\alpha\times\kappa$ in the formalization, but instead use some fresh type 
variable $\nu$ which we only demand to be \emph{isomorphic} to $\alpha\times\kappa$ via the 
morphisms \const{pair\us of\us term} and \const{term\us of\us pair} (this happens once more in a 
locale). So, $\nu$ can be instantiated by $\alpha$ itself if $\kappa$ is instantiated by 
\type{unit}, eventually yielding the ordinary polynomial ring $\polyab$.
\end{remark}


\section{Gr\"obner Bases and Buchberger's Algorithm}
\label{sec::Groebner}

From now on we tacitly assume, unless stated otherwise, that all definitions and theorems are 
stated in context \locale{gd\us term} (meaning that all parameters and axioms of 
\locale{gd\us term} are available for use, that $\kappa$ belongs to type class 
\class{wellorder}, and that $\alpha$ belongs to type class 
\class{graded\us dickson\us powerprod}), 
and that the type $\beta$ of coefficients belongs to type class \class{field}.

\subsection{Polynomial Reduction}
\label{sec::Reduction}

Polynomial reduction is defined analogously to Definition~\ref{def::Reduction}:
\begin{lstlisting}
definition red%\us%single::"$(\polyakb)\Rightarrow(\polyakb)\Rightarrow(\polyakb)\Rightarrow\alpha\Rightarrow$ bool"
  where "red%\us%single p q f t $\longleftrightarrow$ (f $\neq$ 0 $\wedge$ coeff p (t $\circledast$ lt f) $\neq$ 0 $\wedge$
                                  q = p - monom%\us%mult ((coeff p (t $\circledast$ lt f)) / lc f) t f)"

definition red :: "$(\polyakb)$ set $\Rightarrow(\polyakb)\Rightarrow(\polyakb)\Rightarrow$%\,%bool"
  where "red F p q $\longleftrightarrow$ ($\exists$f$\in$F. $\exists$t. red_single p q f t)"
  
definition is%\us%red :: "$(\polyakb)$ set $\Rightarrow$ $(\polyakb)$ $\Rightarrow$ bool"
  where "is%\us%red F a $\longleftrightarrow$ ($\exists$q. red F p q)"
\end{lstlisting}
$\const{red\us single}(p,q,f,t)$ expresses that polynomial $p$ reduces to $q$ modulo the individual 
polynomial $f$, multiplying $f$ by power-product $t$. Likewise, $\const{red}(F)(p,q)$ expresses that 
$p$ reduces to $q$ modulo the set $F$ in 
one step; hence, $\const{red}(F)$ is the actual \emph{reduction relation} modulo $F$, and $(\const{red}(F))^{**}$ denotes its reflexive-transitive closure. In in-line formulas we will use the conventional infix notations $p\rightarrow_F q$ and $p\rightarrow_F^* q$ instead of the more clumsy $\const{red}(F)(p,q)$ and $(\const{red}(F))^{**}(p,q)$, respectively. $\const{is\us red}(F)(p)$, finally, expresses that $p$ is reducible modulo $F$.

\newcommand{\fs}{\ensuremath{f\!s}}
After introducing the above notions, we are able to prove, for instance,
\begin{lstlisting}
lemma red%\us%ord: "red F p q $\Longrightarrow$ q $\ordsp$ p"
\end{lstlisting}
which immediately implies that $\rightarrow_F$ is well-founded. This justifies implementing a 
function \const{trd}, which totally reduces a given vector-polynomial $p$ modulo a finite list $\fs$ 
of polynomials and, thus, computes a \emph{normal form} of $p$ modulo $\fs$. 
Operationally, $\const{trd}(\fs,p)$ iterates over the terms appearing in the polynomial $p$ in 
order (starting with the greatest one) and tries to reduce them modulo the polynomials in the list 
$\fs$; for each term, the first suitable $f\in\fs$\footnote{Abusing notation, $x\in xs$, for a 
list $xs$, means that $x$ is an element of $xs$.} is taken (if any). After implementing \const{trd} 
in said way, it is possible to derive the following characteristic properties:
\begin{lstlisting}
lemma trd%\us%red%\us%rtrancl: "(red (set fs))$^{**}$ p (trd fs p)"

lemma trd%\us%irred: "$\neg$ is%\us%red (set fs) (trd fs p)"
\end{lstlisting}
So, \const{trd} really computes \emph{some} normal form of the given polynomial modulo the given 
list of polynomials. But recall from Section~\ref{sec::Background} that normal forms are in general 
not unique, i.\,e. the reduction relation modulo an arbitrary set $F$ is in general not confluent.

\subsection{Gr\"obner Bases}
\label{sec::GB}

The fact that $\rightarrow_F$ is not confluent for all $F$ motivates the definition of a 
\emph{Gr\"obner basis} as a set that induces a confluent reduction relation:
\begin{lstlisting}
definition is%\us%Groebner%\us%basis :: "($\polyakb$) set $\Rightarrow$ bool"
  where "is%\us%Groebner%\us%basis F $\longleftrightarrow$ is%\us%confluent (red F)"
\end{lstlisting}
where \const{is\us confluent} is the predicate-analogue of \const{CR} from 
\thy{Abstract-Rewriting}~\cite{Abstract-Rewriting-AFP}.

Before we are able to state and prove Theorem~\ref{thm::SPoly}, we need S-polynomials:
\begin{lstlisting}
definition spoly :: "$(\polyakb)\Rightarrow(\polyakb)\Rightarrow(\polyakb)$"
  where "spoly p q = (if snd (lt p) = snd (lt q) then
                        let l = lcm (lp p) (lp q) in
                          (monom_mult (1 / (lc p)) (l / (lp p)) p) -
                          (monom_mult (1 / (lc q)) (l / (lp q)) q)
                      else 0)"
\end{lstlisting}

Theorem~\ref{thm::SPoly} states that a set $F$ is a Gr\"obner basis if the S-polynomials of all 
pairs of elements in $F$ can be reduced to $0$ modulo $F$, i.\,e.
\begin{lstlisting}
theorem Buchberger%\us%thm%\us%finite:
  assumes "finite F"
  assumes "$\bigwedge$p q. p $\in$ F $\Longrightarrow$ q $\in$ F $\Longrightarrow$ (red F)$^{**}$ (spoly p q) 0"
  shows "is%\us%Groebner%\us%basis F"
\end{lstlisting}
The finiteness constraint on $F$ could be weakened to an assumption involving Dickson gradings and 
\const{dgrad\us p\us set}, just as in \thm{ord-p-minimum}. Our proof of 
Theorem~\emph{Buchberger-thm-finite} exploits various results about well-founded binary relations 
formalized in~\cite{Abstract-Rewriting-AFP}. Thanks to that theorem, for deciding whether a finite 
set is a Gr\"obner basis it suffices to compute normal forms of finitely many S-polynomials and 
check whether they are $0$. In fact, also the converse of \emph{Buchberger-thm-finite} holds 
(quite trivially), so whenever one finds a non-zero normal form of an S-polynomial, the given set 
cannot be a Gr\"obner basis.

Another alternative characterization of Gr\"obner bases proved in the formalization is based on the divisibility of leading terms; this characterization
is particularly useful for establishing the correctness of the algorithm for computing reduced Gr\"obner bases in 
Section~\ref{sec::Reduced}:
\begin{lstlisting}
lemma GB%\us%alt%\us%3%\us%finite:
  assumes "finite F"
  shows "is%\us%Groebner%\us%basis F $\longleftrightarrow$
               ($\forall$p$\in$pmdl F. p $\neq$ 0 $\longrightarrow$ ($\exists$f$\in$F. f $\neq$ 0 $\wedge$ lt f $\dvdv$ lt p))"
\end{lstlisting}

\subsection{An Algorithm Schema for Computing Gr\"obner Bases}
\label{sec::Schema}

Theorem~\emph{Buchberger-thm-finite} not only yields an algorithm for deciding whether a given 
finite set $F$ is a Gr\"obner basis or not, but also an algorithm for \emph{completing} $F$ to a 
Gr\"obner basis in case it is not. This algorithm, called \emph{Buchberger's algorithm}, is a 
classical critical-pair/completion algorithm that repeatedly checks whether all S-polynomials reduce 
to $0$, and if not, adds their non-zero normal forms to the basis to make them reducible to $0$; the 
new elements that are added to the basis obviously do not change the submodule generated by the 
basis, since reduction preserves submodule membership.

In our formalization, we do not directly implement Buchberger's algorithm, but instead consider a 
more general algorithm schema first, of which both Buchberger's algorithm and Faug\`ere's $F_4$ 
algorithm (cf. Section~\ref{sec::F4}) are particular instances. This algorithm schema is called 
\const{gb\us schema\us aux} and implemented by the following tail-recursive function:
\begin{lstlisting}
function gb%\us%schema%\us%aux :: "$(\polyakb)$ list $\Rightarrow$
                           $((\polyakb)\times(\polyakb))$ list $\Rightarrow$
                           $(\polyakb)$ list" where
  "gb%\us%schema%\us%aux bs ps =
        (if ps = [] $\vee$ gen%\us%whole%\us%module bs then
          bs
        else
          (let sps = %\param{sel}% bs ps; ps0 = ps -- sps; hs = %\param{compl}% bs ps0 sps in
            gb%\us%schema%\us%aux (bs @ hs) (add%\us%pairs bs ps0 hs)))"
\end{lstlisting}
The first argument of \const{gb\us schema\us aux}, $bs$, is the so-far computed basis, and the 
second argument $ps$ is the list of all pairs of polynomials from $bs$ whose S-polynomials might not 
yet reduce to $0$ modulo $bs$. Hence, as soon as $ps$ is empty all S-polynomials reduce to $0$, and 
by virtue of Theorem~\emph{Buchberger-thm-finite} the list $bs$ constitutes a Gr\"obner basis.
Moreover, if the current basis is detected to generate the whole module,\footnote{The ``whole 
module'' in this context corresponds to $K[x_0,\ldots,x_n]^k$, where $x_0,\ldots,x_n$ are the 
indeterminates and $k$ is the largest component-index appearing in $bs$.} then it 
can as well be returned immediately without any further ado. \const{gen\us whole\us module} 
basically checks whether for each of the finitely many component-indices $i$ appearing in $bs$ there 
is a non-zero polynomial $b$ in $bs$ such that the component-index of $\const{lt}(b)$ is $i$ and 
$\const{lp}(b)=1$. This parallels the scalar case, where a set of polynomials is known to generate 
the whole ring if it contains a non-zero constant polynomial.

The auxiliary function \const{add\us pairs}, when applied to arguments $bs$, $ps_0$ and $hs$, 
returns a new list of pairs of polynomials which contains precisely (i)~all pairs from $ps_0$, 
(ii)~the pair $(h,b)$ for all $h\in hs$ and $b\in bs$, and (iii)~one of the pairs $(h_1,h_2)$ or 
$(h_2,h_1)$ for all $h_1,h_2\in hs$ with $h_1\neq h_2$. The auxiliary function \const{diff\us list} 
(infix ``\const{-{}-}'') is the analogue of set-difference for lists, i.\,e. it removes all 
occurrences of all elements of its second argument from its first argument. \const{append}, infix 
\const{@}, concatenates two lists.

The two functions \param{sel} and \param{compl} are additional parameters of the algorithm; they 
are not listed among the arguments of \const{gb\us schema\us aux} here merely for the sake of 
better readability. Informally, they are expected to behave as follows:
\begin{itemize}
 \item If $ps$ is non-empty, $\param{sel}(bs,ps)$ should return a non-empty sublist $sps$ of $ps$.
 
 \item $\param{compl}(bs,ps,sps)$ should return a (possibly empty) list $hs$ of polynomials such that (i)~$0\notin hs$, (ii)~$hs\subseteq\const{pmdl}(bs)$, (iii)~$\const{spoly}(p,q)\rightarrow_{bs\cup hs}^* 0$ for all $(p,q)\in sps$, and (iv)~$\neg\,\const{lt}(b)\ \dvdv\ \const{lt}(h)$ for all $b\in bs$ and $h\in hs$.
\end{itemize}
Typically, concrete instances of \param{sel} do not take $bs$ into account, but in any case it 
does not harm to pass it as an additional argument. Any instances of the two parameters that 
satisfy the above requirements lead to a partially correct procedure for computing Gr\"obner bases, 
since \param{compl} takes care that all S-polynomials of the selected pairs $sps$ reduce to $0$. 
However, the procedure is not only partially correct, but also terminates for every input; the 
argument roughly proceeds as follows:
\begin{itemize}
 \item Assume the procedure did not terminate. Then, infinitely many non-zero polynomials $h$ (originating from \param{compl}) are added to the basis $bs$.
 
 \item The leading term of each of these polynomials is not divisible (w.\,r.\,t. $\dvdv$) by the leading term of any polynomial in the current basis.

 \item Therefore, the sequence of these polynomials violates the \emph{Dickson property} of 
sequences of power-products -- a contradiction. This argument also works if $\kappa$ is 
instantiated by \type{nat} and $\alpha$ by $\polym{\type{nat}}{\type{nat}}$, because the sets 
of indeterminates and non-zero components appearing in the current basis $bs$ in recursive calls of 
\const{gb\us schema\us aux} are finite and uniformly bounded.
\end{itemize}
Function \const{gb\us schema}, finally, calls \const{gb\us schema\us aux} with the right initial values:
\begin{lstlisting}
definition gb%\us%schema :: "($\polyakb$) list $\Rightarrow$ ($\polyakb$) list"
  where "gb%\us%schema bs = gb%\us%schema%\us%aux bs (add%\us%pairs [] [] bs)"
\end{lstlisting}

\begin{remark}
Not only Buchberger's algorithm and $F_4$ are instances of \const{gb\us schema}, as briefly indicated above and discussed in more detail in Sections~\ref{sec::BB} and~\ref{sec::F4}, but also Faug\`ere's $F_5$ algorithm~\cite{Faugere02} is an instance of it. $F_5$ is currently the most efficient method for computing Gr\"obner bases, but formalizing it in a proof assistant is a challenging task that is yet to be accomplished. $F_5$ computes Gr\"obner bases \emph{incrementally}, i.\,e. for computing a Gr\"obner basis of $m$ polynomials it calls itself recursively on the first $m-1$ polynomials and then adds the $m$-th polynomial. \const{gb\us schema} can handle such incremental computations as well, although this is not reflected in its (simplified) presentation in this paper.
\end{remark}

\subsection{Buchberger's Algorithm}
\label{sec::BB}

The function implementing the usual Buchberger algorithm, called \const{gb}, can immediately be obtained from \const{gb\us schema} by instantiating
\begin{itemize}
 \item \param{sel} to a function that selects a single pair, i.\,e. returns a singleton list, and
 \item \param{compl} to a function that totally reduces $\const{spoly}(p,q)$ to some normal form $h$ using \const{trd}, where $(p,q)$ is the pair selected by the instance of \param{sel}, and returns the singleton list $[h]$ if $h\neq 0$ and the empty list otherwise.
\end{itemize}
These instances of \param{sel} and \param{compl} can easily be proved to meet the requirements listed above, so we can finally conclude that \const{gb} indeed always computes a Gr\"obner basis of the submodule generated by its input:
\begin{lstlisting}
theorem gb%\us%isGB: is%\us%Groebner%\us%basis (set (gb bs))

theorem gb%\us%pmdl: pmdl (set (gb bs)) = pmdl (set bs)
\end{lstlisting}

Gr\"obner bases have many interesting properties. One of them was briefly sketched at the end of 
Section~\ref{sec::Background}: if $G$ is a Gr\"obner basis, then a polynomial $p$ is in the 
submodule generated by $G$ iff the unique normal form of $p$ modulo $G$ is $0$. Together with the 
two previous theorems this observation leads to an effective answer to the membership problem for 
submodules represented by finite lists of generators:
\begin{lstlisting}
theorem in%\us%pmdl%\us%gb: "p $\in$ pmdl (set bs) $\longleftrightarrow$ (trd (gb bs) p) = 0"
\end{lstlisting}

\subsection{Improving Efficiency: Buchberger's Criteria}
\label{sec::Criteria}

The key ingredient of \const{gb\us schema\us aux} is parameter \param{compl}, since it is precisely this function that has to ensure that all S-polynomials can be reduced to $0$ modulo the enlarged list $bs @ hs$. However, computing the new polynomials $hs$ can be very time-consuming, since it is usually accomplished by reducing the S-polynomials to some normal form, either employing \const{trd} or some other methodology (as in $F_4$). Now, if the S-polynomial of some pair $(p,q)$ can be reduced to $0$ modulo the \emph{current} basis $bs$ already, and this fact can somehow be predicted without actually \emph{doing} the reduction, the whole expensive normal-form computation of $\const{spoly}(p,q)$ could be avoided altogether; in this case, $(p,q)$ is called a \emph{useless pair}. Therefore, one approach to improve the efficiency of \const{gb\us schema\us aux} is to detect as many useless pairs as possible and to immediately discard them without passing them along to the normal-form computation. Since all this 
happens in \param{compl}, and \param{compl} is a parameter of \const{gb\us schema\us aux}, many different strategies for detecting useless pairs can be implemented easily, without having to change the overall implementation of the algorithm schema.

In our formalization, the two instantiations of \param{compl} yielding Buchberger's algorithm and 
the $F_4$ algorithm (cf. Section~\ref{sec::F4}), respectively, incorporate two standard criteria, 
originally due to Buchberger, for detecting useless pairs: the \emph{product criterion} (which is 
only applicable in the scalar case with $\kappa=\type{unit}$) and the \emph{chain criterion}. In a 
nutshell, 
the product criterion states that if $\const{gcd}(\const{lp}(p),\const{lp}(q))=1$, then $\const{spoly}(p,q)\rightarrow_{\{p,q\}}^*0$. The chain criterion states that if there is some 
$r$ in the current basis $bs$ satisfying (i)~$\const{lp}(r)\ \const{dvd}\ \const{lcm}(\const{lp}(p),\const{lp}(q))$, (ii)~$\const{spoly}(p,r)\rightarrow_{bs}^* 0$ and (iii)~$\const{spoly}(r,q)\rightarrow_{bs}^* 0$, then also 
$\const{spoly}(p,q)\rightarrow_{bs}^* 0$, and hence the pair $(p,q)$ can be discarded. A more thorough account 
on Buchberger's criteria can be found in~\cite{Buchberger79}. In the 
formalization we of course prove that these two criteria are indeed correct, in the sense that they 
only discard \emph{useless} pairs. It must be noted, though, that in general neither of them detects 
\emph{all} useless pairs.

\begin{remark}
When testing the chain criterion on concrete input, lots of equality-checks between polynomials have to be performed. Because of that, \const{gb\us schema} automatically assigns a unique ID (of type \type{nat}) to every polynomial, allowing the chain criterion to only compare IDs rather than full polynomials. This small trick helped to significantly increase the efficiency of the implementation.
\end{remark}

\section{Faug\`ere's $F_4$ Algorithm}
\label{sec::F4}

In Buchberger's algorithm, in each iteration precisely \emph{one} S-polynomial is reduced modulo 
the current basis, giving rise to at most \emph{one} new basis element. However, as J.-C. Faug\`ere 
observed in~\cite{Faugere99}, it is possible to reduce several S-polynomials \emph{simultaneously} 
with a considerable gain of efficiency (especially for \emph{large} input). To that end, one selects 
some pairs from the list $ps$, reduces them modulo the current basis, and adds the resulting 
non-zero normal forms to the basis; in short, several iterations of the usual Buchberger algorithm 
are combined into one single iteration. This new algorithm is called $F_4$.

The crucial idea behind $F_4$, and the reason why it can be much faster than 
Buchberger's algorithm, is the clever implementation of simultaneous reduction 
by computing the \emph{reduced row echelon form} of certain coefficient 
matrices. Before we can explain how this works, we need a couple of 
definitions; let $\fs$ always be a list of vector-polynomials of length $m$, and $vs$ 
be a list of terms of length $\ell$.
\begin{itemize}
 \item $\const{Keys\us to\us list}(\fs)$ returns the list of all distinct terms appearing in $\fs$, 
sorted \emph{descending} w.\,r.\,t. $\ordv$.
 
 \item $\const{polys\us to\us mat}(vs,\fs)$ returns a matrix $A$ (\`a 
la~\cite{Jordan_Normal_Form-AFP}) of dimension $m\times \ell$, satisfying 
 $A_{i,j}=\C(\fs_i,vs_j)$, for $0\leq i<m$ and $0\leq j<\ell$.\footnote{We use $0$-based indexing 
of lists, vectors and matrices, just as Isabelle/HOL.}
 
 \item $\const{mat\us to\us polys}(vs,A)$ is the ``inverse'' of \const{polys\us 
 to\us mat}, i.\,e. if $A$ is a matrix of dimension $m\times \ell$ it returns the 
 list $gs$ of polynomials satisfying 
 $\C(gs_i,vs_j)=A_{i,j}$ and $\C(gs_i,v)=0$ for 
 all other terms $v$ not contained in $vs$.
 
 \item $\const{row\us echelon}(A)$ returns the reduced row echelon form of matrix $A$; it is defined in terms of \const{gauss\us jordan} from~\cite{Jordan_Normal_Form-AFP}.
\end{itemize}
With these auxiliary functions at our disposal we can now give the formal definitions of two concepts whose importance will become clear below:
\begin{lstlisting}
definition Macaulay%\us%mat :: "($\polyakb$) list $\Rightarrow\beta$::field mat"
  where "Macaulay%\us%mat fs = polys%\us%to%\us%mat (Keys%\us%to%\us%list fs) fs"

definition Macaulay%\us%red :: "($\polyakb$) list $\Rightarrow$ ($\polyakb$::field) list"
  where "Macaulay%\us%red fs =
     (let lts = map lt (filter ($\lambda$f. f $\neq$ 0) fs) in
       filter ($\lambda$f. f $\neq$ 0 $\wedge$ lt f $\notin$ set lts)
              (mat%\us%to%\us%polys (Keys%\us%to%\us%list fs) (row%\us%echelon (Macaulay%\us%mat fs))))"
\end{lstlisting}
$\const{Macaulay\us mat}(\fs)$ is called the \emph{Macaulay matrix} of $\fs$. 
$\const{Macaulay\us red}(\fs)$ constructs the Macaulay matrix of $\fs$, 
transforms it into reduced row echelon form, and converts the resulting matrix 
back to a list of polynomials, from which it filters out those non-zero 
polynomials whose leading terms are not among the leading terms of the original list $\fs$.
\begin{example}
Let $\fs$ be the list $\fs=[x_1^3-5x_0^2x_1-2,-4x_1^3+2x_1^2+x_0^2x_1,2x_1^3-x_1^2-x_0+4]$ of 
scalar polynomials, and 
let $\preceq$ and $\ordv$ be the lexicographic order with $x_0\prec x_1$. The sorted list of terms 
(or, 
in this case, power-products) appearing in $\fs$ is $[x^3,x^2,xy^2,y,1]$. Hence, the Macaulay 
matrix of $\fs$ is
\[
\bordermatrix{
~ & x_1^3 & x_1^2 & x_0^2x_1 & x_0 & 1\cr
~ & 1 &  0 & -5 &  0 & -2\cr
~ & -4 &  2 &  1 &  0 &  0\cr
~ & 2 & -1 &  0 & -1 &  4
}
\]
Row-reducing the Macaulay matrix yields
\[
\bordermatrix{
~ & x_1^3 & x_1^2 & x_0^2x_1 & x_0 & 1\cr
~ & 1 & 0 & 0 & -10 & 38\cr
~ & 0 & 1 & 0 & -19 & 72\cr
~ & 0 & 0 & 1 &  -2 &  8
}
\]
from which we can extract the three polynomials $h_1=x_1^3-10x_0+38$, 
$h_2=x_1^2-19x_0+72$ and $h_3=x_0^2x_1-2x_0+8$. The leading term 
of $h_1$ is $x_1^3$, which is also the leading term of one of the three (actually all three) 
original polynomials. So, 
$\const{Macaulay\us red}(\fs)$ returns the two-element list $[h_2,h_3]$.
\end{example}

\const{Macaulay\us red} is the key ingredient of the $F_4$ algorithm, because 
the list it returns is precisely the list $hs$ that must be 
added to $bs$ in an iteration of \const{gb\us schema\us aux}. The only question 
that still remains open is which argument list $\fs$ it needs to be applied to; 
this question is answered by an algorithm called \emph{symbolic preprocessing}, 
implemented by function \const{sym\us preproc} in our formalization. 
\const{sym\us preproc} takes two arguments, namely the current basis $bs$ and 
the list of selected pairs $sps$, and informally behaves as follows:
\begin{enumerate}
 \item For each $(p,q)\in sps$ compute the two polynomials
 \[
 \const{monom\us mult}(1/\lc(p),\const{lcm}(\lp(p),\lp(q))/\lp(p),p)
 \]
 \[
 \const{monom\us mult}(1/\lc(p),\const{lcm}(\lp(p),\lp(q))/\lp(p),p)
 \]
 whose difference is precisely $\spoly(p,q)$ (unless $\const{spoly}(p,q)$ is $0$). Collect all 
 these polynomials in an auxiliary list $\fs'$.
  
 \item Collect all monomial multiples of each $b\in bs$ that are needed to 
 totally reduce the elements of $\fs'$ in a list $\fs''$. That means, for all 
 $b,f,g,h$ with $b\in bs$, $f\in \fs'$, $f\rightarrow_{bs}^* g$ and 
 $\const{red\us single}(g,h,b,t)$, the monomial multiple $\const{monom\us 
 mult}(1,t,b)$ of $b$ must be included in $\fs''$.
 
 \item Return the concatenation $\fs' @ \fs''$ of $\fs'$ and $\fs''$.
\end{enumerate}
The interesting part of symbolic preprocessing is Step~2, which can be accomplished without actually carrying out the reductions. In our formalization, it is implemented by the two functions \const{sym\us preproc\us addnew} and \const{sym\us preproc\us aux}, defined as
\begin{lstlisting}
primrec sym%\us%preproc%\us%addnew :: "($\polyakb$) list $\Rightarrow$ $(\alpha\times\kappa)$ list $\Rightarrow$
                               ($\polyakb$) list $\Rightarrow$ $(\alpha\times\kappa)$ $\Rightarrow$
                               ($(\alpha\times\kappa)$ list $\times$ ($\polyakb$) list)" where
  "sym%\us%preproc%\us%addnew [] vs fs _ = (vs, fs)"|
  "sym%\us%preproc%\us%addnew (b # bs) vs fs v =
    (if (lt b) $\dvdv$ v then
      let f = monom%\us%mult 1 (fst v / (lp b)) b in
        sym%\us%preproc%\us%addnew bs (merge%\us%wrt (op $\succ_{\sf v}$) vs (keys%\us%to%\us%list (tail f)))
                           (insert%\us%list f fs) v
    else
      sym%\us%preproc%\us%addnew bs vs fs v)"

function sym%\us%preproc%\us%aux :: "($\polyakb$) list $\Rightarrow$
                             ($(\alpha\times\kappa)$ list $\times$ ($\polyakb$) list) $\Rightarrow$
                             ($\polyakb$) list" where
  "sym%\us%preproc%\us%aux bs (vs, fs) =
    (if vs = [] then
      fs
    else
      let v = Max (set vs); vs' = removeAll v vs in
        sym%\us%preproc%\us%aux bs (sym%\us%preproc%\us%addnew bs vs' fs v))"
\end{lstlisting}
\const{sym\us preproc} calls \const{sym\us preproc\us aux} with the current 
basis $bs$ and the pair $(vs,\fs')$, where $\fs'$ is the result of Step~1 and 
$vs$ is the sorted list of terms appearing in $\fs'$. A more detailed 
account on symbolic preprocessing can be found in~\cite{Faugere99}.

Putting everything together, the function \const{f4\us red} is obtained as
\begin{lstlisting}
definition f4%\us%red::"($\polyakb$::field) list $\Rightarrow$
                    ($\polyakb$ $\times$ $\polyakb$) list $\Rightarrow$ ($\polyakb$) list"
  where "f4%\us%red bs sps = Macaulay%\us%red (sym%\us%preproc bs sps)"
\end{lstlisting}
and proved to be a feasible instance of parameter \param{compl};\footnote{Actually, the instance is not exactly \const{f4\us red}, but a function that also discards useless pairs according to Section~\ref{sec::Criteria}.} in particular, the leading terms of the polynomials in $hs=\const{f4\us red}(bs,sps)$ are not divisible by the leading terms of the polynomials in $bs$, and indeed all S-polynomials originating from pairs in $sps$ are reducible to $0$ modulo the enlarged basis $bs @ hs$:
\begin{lstlisting}
lemma f4%\us%red%\us%not%\us%dvd:
  assumes "h $\in$ set (f4%\us%red bs sps)" and "b $\in$ set bs" and "b $\neq$ 0"
  shows "$\neg$ lt b $\dvdv$ lt h"

lemma f4%\us%red%\us%spoly%\us%reducible:
  assumes "set sps $\subseteq$ set bs $\times$ set bs" and "(p, q) $\in$ set sps"
  shows "(red (set (bs @ (f4%\us%red bs sps))))$^{**}$ (spoly p q) 0"
\end{lstlisting}
Eventually, the resulting instance of \const{gb\us schema} which implements Faug\`ere's $F_4$ algorithm is called \const{f4}.

Summarizing, the simultaneous reduction of several S-polynomials boils down to the computation of the reduced row echelon form of Macaulay matrices over the coefficient field $K$. These matrices are typically very big, very rectangular (i.\,e. have much more columns than rows) and extremely sparse. Therefore, if $F_4$ is to outperform Buchberger's algorithm, such matrices must be stored efficiently (possibly even involving some sort of \emph{compression}), and the computation of reduced row echelon forms must be highly optimized; we again refer to~\cite{Faugere99} for more information. Furthermore, the superiority of $F_4$ over Buchberger's original algorithm only takes effect when the problem instances are sufficiently large as to outweigh the overhead stemming from all the matrix constructions in $F_4$; for small problems ($\lesssim 5$ indeterminates, moderate degrees), Buchberger's algorithm is typically still faster.

\section{Reduced Gr\"obner Bases}
\label{sec::Reduced}

As mentioned in Section~\ref{sec::Background}, Gr\"obner bases are not unique, even if the term 
order $\ordv$ is fixed. One can, however, impose stronger constraints on generating sets of 
submodules than giving rise to a confluent reduction relation, which \emph{do} ensure uniqueness; 
the resulting concept is that of \emph{reduced Gr\"obner bases}.

The central idea behind reduced Gr\"obner bases is \emph{auto-reducedness}: a set $B$ of 
polynomials is auto-reduced iff no $b\in B$ can be reduced modulo $B\backslash\{b\}$. A 
reduced Gr\"obner basis, then, is simply an auto-reduced Gr\"obner basis of non-zero 
monic\footnote{A polynomial is called \emph{monic} iff its leading coefficient is $1$.}
polynomials:
\begin{lstlisting}
definition is%\us%reduced%\us%GB :: "($\polyakb$) set $\Rightarrow$ bool"
  where "is%\us%reduced%\us%GB B $\longleftrightarrow$ is%\us%Groebner%\us%basis B $\wedge$ is%\us%auto%\us%reduced B $\wedge$
                              is%\us%monic%\us%set B $\wedge$ 0 $\notin$ B"
\end{lstlisting}

After having defined reduced Gr\"obner bases as above, one can prove with moderate effort that, upon existence, they are indeed unique for every submodule (of course only modulo the implicitly fixed ordering $\ordv$):
\begin{lstlisting}
theorem is%\us%reduced%\us%GB%\us%unique:
  assumes "is%\us%reduced%\us%GB A" and "is%\us%reduced%\us%GB B" and "pmdl A = pmdl B"
  shows "A = B"
\end{lstlisting}

Besides uniqueness, one can furthermore also prove \emph{existence} of reduced Gr\"obner bases. The proof we give in the formalization is even constructive, in the sense that we formulate an algorithm which auto-reduces and makes monic a given set (or, more precisely, list) of polynomials, and, therefore, when applied to \emph{some} Gr\"obner basis returns \emph{the reduced} Gr\"obner basis of the submodule it generates. Said algorithm proceeds in three steps:
\begin{enumerate}
 \item First, all polynomials of the input list whose leading terms are divisible by the leading term of any other polynomial in the input, are removed, and so are all occurrences of $0$.
 
 \item Next, every remaining polynomial is totally reduced modulo the others (employing function \const{trd}) and replaced by the result of this process.
 
 \item Finally, the polynomials are made monic by dividing through their respective leading coefficients.
\end{enumerate}
The function combining these three steps is called \const{comp\us red\us monic\us basis} and possesses the following two key properties:
\begin{lstlisting}
lemma comp%\us%red%\us%monic%\us%basis%\us%is%\us%reduced%\us%GB:
  assumes "is%\us%Groebner%\us%basis (set bs)"
  shows "is%\us%reduced%\us%GB (set (comp%\us%red%\us%monic%\us%basis bs))"

lemma comp%\us%red%\us%monic%\us%basis%\us%pmdl:
  assumes "is%\us%Groebner%\us%basis (set bs)"
  shows "pmdl (set (comp%\us%red%\us%monic%\us%basis bs)) = pmdl (set bs)"
\end{lstlisting}
So, by combining functions \const{gb} (or \const{f4}) and \const{comp\us red\us monic\us basis}, we 
obtain a certified function for computing reduced Gr\"obner bases from any given list of 
polynomials, and can moreover conclude that every finitely-generated submodule of $K[X]^k$ has a 
unique reduced Gr\"obner basis:
\begin{lstlisting}
theorem exists%\us%unique%\us%reduced%\us%GB%\us%finite:
  assumes "finite F"
  shows "$\exists!$G. is%\us%reduced%\us%GB G $\wedge$ pmdl G = pmdl F"
\end{lstlisting}

\begin{example}
The Gr\"obner basis $\tilde{F}=\{x_1^2,x_0x_1+x_0^2,x_0^3\}$ computed in Example~\ref{ex::GB} is 
already the reduced Gr\"obner basis of the ideal it generates. On the other hand, 
$G=\{x_0^3-2x_0x_1+1,x_1-x_0\}$ is a Gr\"obner basis (w.\,r.\,t. the degree-lexicographic ordering 
with $x_0\prec x_1$) but no reduced Gr\"obner basis, because the first element is reducible modulo 
the second one.
\end{example}

\begin{remark}
Auto-reduction can already be applied during the computation of Gr\"obner bases and may lead to a significant speed-up of the algorithm. Incorporating auto-reduction into \const{gb\us schema} is possible future work.
\end{remark}


\section{Gr\"obner Bases of Syzygy Modules}
\label{sec::Syzygy}

Given a list $bs=[b_0,\ldots,b_{m-1}]$ of $m$ vector-polynomials, one could ask oneself 
what the polynomial 
relations among the elements of $bs$ are. In other words, one wants to find all $m$-component 
vectors $s=(s_0,\ldots,s_{m-1})^{\mathrm{T}}\in K[X]^m$ such that
\[\sum_{i=0}^{m-1}s_i\,b_i=0.\] 
In the 
literature, such a vector $s$ of polynomials is called a \emph{syzygy}~\cite{Kreuzer00} of $bs$, 
and as one can easily see the set of all syzygies of a list $bs$ forms a submodule of $K[X]^m$.
\begin{example}
If $m=2$ and $b_0,b_1\in K[X]\backslash\{0\}$, 
then a non-trivial syzygy is obviously given by
$
\left(\begin{array}{c}b_1\\-b_0\end{array}\right)
$,
because $b_1\,b_0+(-b_0)\,b_1=0$.
More generally, each list of scalar polynomials with at least two non-zero elements admits 
non-trivial syzygies of the above kind.
\end{example}

As it turns out, it is not difficult to compute Gr\"obner bases of syzygy modules. We briefly 
outline how it works in theory; so, assume that $bs$ is the $m$-element list $[b_0,\ldots,b_{m-1}]$ 
of 
polynomials in $K[X]^k$.\footnote{In the formalization the type of the polynomials is 
$\polym{(\polym{\chi}{\type{nat}})}{\beta}$, because fixing $\kappa$ to \type{nat} turns out to be 
convenient here.}
\begin{enumerate}
 \item Add further components to the $b_i$, for all $0\leq i<m$, such that $b_i$ becomes 
$(0,\ldots,0,1,0,\ldots,0,b_i)^{\mathrm{T}}$, where the $1$ occurs precisely in the $i$-th 
component. Note: these vectors have $\geq m+1$ components, since after the newly introduced $m$ 
components come \emph{all} of $b_i$'s existing components.
Call the resulting list $bs'$.
 
 \item Compute a Gr\"obner basis $gs$ of $bs'$ w.\,r.\,t. a POT-extension $\ordv$ of some 
admissible order $\preceq$ on power-products.
 
 \item From $gs$ extract those elements of the form $(s_0,\ldots,s_{m-1},0,\ldots,0)^{\mathrm{T}}$, 
and restrict them to $m$-dimensional vectors $(s_0,\ldots,s_{m-1})^{\mathrm{T}}$. These vectors 
constitute a Gr\"obner 
basis w.\,r.\,t. $\ordv$ of the syzygy module of $bs$.
\end{enumerate}

\begin{example}
\label{ex::syzygy}
Consider the three scalar polynomials $b_0=x_0x_1-x_2$, $b_1=x_0x_2-x_1$ and $b_2=x_1x_2-x_0$, and 
the list $bs=[b_0,b_1,b_2]$. According to Step~1 we construct $bs'$ as
\[
bs'=[
\left(\begin{array}{c}
1\\0\\0\\b_0
\end{array}\right),
\left(\begin{array}{c}
0\\1\\0\\b_1
\end{array}\right),
\left(\begin{array}{c}
0\\0\\1\\b_2
\end{array}\right)
].
\]
Next, we compute a (non-reduced) Gr\"obner basis $gs$ of $bs'$ w.\,r.\,t. the POT-extension of the 
degree-reverse-lexicographic order with $x_0\prec x_1\prec x_2$:
\begin{eqnarray}
gs & = & [
\left(\begin{array}{c}
1\\0\\0\\b_0
\end{array}\right),
\left(\begin{array}{c}
0\\1\\0\\b_1
\end{array}\right),
\left(\begin{array}{c}
0\\0\\1\\b_2
\end{array}\right),
\left(\begin{array}{c}
-b_1\\b_0\\0\\0
\end{array}\right),
\left(\begin{array}{c}
-b_2\\0\\b_0\\0
\end{array}\right),
\left(\begin{array}{c}
0\\-b_2\\b_1\\0
\end{array}\right),\notag\\
& &\quad
\left(\begin{array}{c}
0\\-x_1\\x_0\\x_1^2-x_0^2
\end{array}\right),
\left(\begin{array}{c}
x_2\\0\\-x_0\\x_0^2-x_2^2
\end{array}\right),
\left(\begin{array}{c}
x_1-x_1x_2^2\\x_1^2x_2-x_2\\x_1^2-x_2^2\\0
\end{array}\right),
\left(\begin{array}{c}
-x_1\\-x_0x_1\\x_0^2-1\\x_0-x_0^3\notag
\end{array}\right)
].
\end{eqnarray}
So, according to Step~3 the four-element list
\[
syz=[
\left(\begin{array}{c}
-b_1\\b_0\\0
\end{array}\right),
\left(\begin{array}{c}
-b_2\\0\\b_0
\end{array}\right),
\left(\begin{array}{c}
0\\-b_2\\b_1
\end{array}\right),
\left(\begin{array}{c}
x_1-x_1x_2^2\\x_1^2x_2-x_2\\x_1^2-x_2^2
\end{array}\right)
]
\]
constitutes a Gr\"obner basis of the syzygy module of $bs$.
\end{example}

In the formalization, function \const{init\us syzygy\us list} takes care of the first step. The 
second step can of course be accomplished by any function for computing Gr\"obner bases, like 
\const{gb} or \const{f4}; there is nothing special about it concerning syzygies. The last step, 
finally, is implemented by function \const{filter\us syzygy\us basis}:
\begin{lstlisting}
definition filter%\us%syzygy%\us%basis :: "nat $\Rightarrow$ $\polym{(\alpha\times\type{nat})}{\beta}$ list $\Rightarrow$ $\polym{(\alpha\times\type{nat})}{\beta}$ list"
  where "filter%\us%syzygy%\us%basis m gs = [g$\leftarrow$gs. snd %\textasciigrave% keys g $\subseteq$ %\{%0..<m%\}%]"
\end{lstlisting}

The correctness of the three-step algorithm sketched above is established by the following 
two lemmas:
\begin{lstlisting}
lemma pmdl%\us%filter%\us%syzygy%\us%basis:
  assumes "distinct bs" and "is%\us%Groebner%\us%basis (set gs)" and
    "pmdl (set gs) = pmdl (set (init%\us%syzygy%\us%list bs))"
  shows "pmdl (set (filter%\us%syzygy%\us%basis (length bs) gs)) = syzygy%\us%module%\us%list bs"

lemma filter%\us%syzygy%\us%basis%\us%isGB:
  assumes "distinct bs" and "is%\us%Groebner%\us%basis (set gs)"
    and "pmdl (set gs) = pmdl (set (init%\us%syzygy%\us%list bs))"
  shows "is%\us%Groebner%\us%basis (set (filter%\us%syzygy%\us%basis (length bs) gs))"
\end{lstlisting}
\const{syzygy\us module\us list} is an auxiliary constant that gives the set of syzygies of its 
argument. The first lemma states that the result of the algorithm indeed generates the syzygy 
module of the input list $bs$, and the second lemma states that the result is even a Gr\"obner 
basis. Neither of the lemmas is confined to a particular function for computing the Gr\"obner basis 
$gs$.

Besides a Gr\"obner basis of the syzygy module of $bs$, the Gr\"obner basis $gs$ computed in Step~2 
carries further useful information:
\begin{itemize}
 \item Projecting $gs$ onto the last component(s), i.\,e. removing those components that were added 
in Step~1, yields a Gr\"obner basis of the original list $bs$.

 \item Each element in $gs$ possesses the property that its first $m$ components are the cofactors, 
w.\,r.\,t. $bs$, of its last component(s).
\end{itemize}
Both these claims are proved in the formalization.
\begin{example}
Continuing Example~\ref{ex::syzygy} we find that
\[
[b_0,b_1,b_2, x_1^2-x_0^2,x_0^2-x_2^2,x_0-x_0^3]
\]
constitutes a Gr\"obner basis of $bs$, and that
\begin{eqnarray}
x_1^2-x_0^2 & = & -x_1\,b_1+x_0\,b_2\notag\\
x_0^2-x_2^2 & = & x_2\,b_0-x_0\,b_2\notag\\
x_0-x_0^3 & = & -x_1\,b_0-x_0x_1\,b_1+(x_0^2-1)\,b_2.\notag
\end{eqnarray}
\end{example}


\section{Code Generation}
\label{sec::Code}

The algorithms about (reduced) Gr\"obner bases and syzygies formalized in Isabelle/HOL can be 
turned into actual Haskell/OCaml/Scala/SML code by means of Isabelle's \emph{Code 
Generator}~\cite{CodeGen}. The formalization contains a couple of sample computations, both by 
Buchberger's algorithm and by the $F_4$ algorithm. One of the main reasons why we deem 
our formalization elegant is that setting up a computation is particularly easy: one does not have 
to care about the (number of) indeterminates featuring in a computation, as indeterminates are 
simply indexed by natural numbers, and it does not matter how many of them appear in a computation; 
the same is true for the dimension $k$ of the module under consideration, since in any case $\kappa$ 
can just be instantiated by \type{nat}. In particular, it is not necessary to a-priori introduce 
dedicated types for univariate/bivariate/trivariate/\ldots\ polynomials.

Table~\ref{tab::Timings} shows the performance of the code generated from our certified algorithms 
on some benchmark problems, on a standard desktop computer. For comparison, Th\'ery~\cite{Thery2001} 
reports $2$~seconds for Cyclic$_5$ and $30$~minutes for Cyclic$_6$, which on a our computer 
reduces to $0.5$~seconds and $2$~minutes, respectively. We identified two main sources of 
inefficiency in our implementation, which explain the huge performance differences:

\begin{itemize}
 \item In computations, power-products and polynomials are represented as associative lists. This 
representation is feasible but not optimal, since certain \emph{invariants} could additionally be 
encoded in the representing type and exploited in code equations. For instance, one could require 
the keys in associative lists representing polynomials to be distinct and sorted descending 
w.\,r.\,t. the chosen ordering $\ordv$; then, most basic operations (addition, subtraction, 
\const{lt}, etc.) could be implemented much more efficiently than is currently the 
case.\footnote{This is also proposed in~\cite{Haftmann2014}.} Although doable in principle, this is 
challenging since Isabelle does not support dependent types, which makes it difficult to encode 
parametric invariants in types (parametric because arbitrary term orders shall be supported). 
Improving the representation of power-products and polynomials is ongoing work, but first 
experiments indicate that such an improvement would indeed reduce computation times drastically: 
Buchberger's algorithm only takes $40$ minutes for Cyclic$_6$ then.

 \item Computations of Gr\"obner bases over $\mathbb{Q}$ often suffer from a so-called 
\emph{coefficient swell}, i.\,e. the numerators and denominators of coefficients grow extremely 
large. For instance, in Cyclic$_6$ the largest denominators occurring during the computation have 
more than $200$ digits. Hence, in order to handle such big numbers efficiently, one should use 
native types of rational numbers in the target languages (like \type{Ratio.ratio} in 
OCaml, which is used in~\cite{Thery2001}). Isabelle's Code Generator, however, constructs its own 
type of rational numbers as pairs of (native) integers. Experiments show that this setup is slower 
by a factor of about $20$ compared to OCaml's \type{Ratio.ratio} when adding rational numbers with 
$200$-digit denominators.
\end{itemize}

So, we have good reason to believe that Cyclic$_6$ can be solved by our certified implementation of Buchberger's 
algorithm in roughly the same amount of time as by Th\'ery's implementation in Coq+OCaml, once the 
two issues listed above have been sorted out. Still, the computation times 
are much slower than in state-of-the-art computer algebra systems like Maple~\cite{Maple} or
Mathematica~\cite{Mathematica}, which return the reduced Gr\"obner basis of Cyclic$_6$ in a split 
second.

\setlength{\tabcolsep}{4pt}
\begin{table}[t]
\center
\begin{tabular}{l | r | r | r | r | r | r}
 & Cyclic$_4$ & Cyclic$_5$ & Cyclic$_6$ & Katsura$_3$ & Katsura$_4$ & 
Katsura$_5$\\\hline
Buchberger & $<1$ & $80$ & $>3600$ & $2$ & $90$ & $>3600$\\
$F_4$ & $2$ & $1900$ & $>3600$ & $40$ & $>3600$ & $>3600$
\end{tabular}
\caption{Timings on benchmark problems, in seconds.}
\label{tab::Timings}
\end{table}

Moreover, our implementation of the $F_4$ algorithm is in most cases considerably slower than 
Buchberger's algorithm, although in Section~\ref{sec::F4} we claimed that it should be the other 
way 
round. One reason for this phenomenon certainly lies in the dense representation of matrices 
in~\cite{Jordan_Normal_Form-AFP} as IArrays of IArrays. As indicated in Section~\ref{sec::F4}, a 
much more efficient representation of matrices in conjunction with optimized algorithms for 
computing reduced row echelon forms would be necessary to outperform Buchberger's algorithm. Such 
representations and algorithms have not been formalized in Isabelle yet. Indeed, our motivation 
to consider $F_4$ was mainly academic interest, to demonstrate it can be formalized with moderate 
effort if only the algorithm schema \const{gb\us schema} is formulated in a sufficiently general 
way, to establish the connection between the computation of Gr\"obner bases and Macaulay matrices, 
and to lay the foundation for more efficient Gr\"obner-basis computations by computer-certified 
state-of-the-art algorithms (like $F_5$~\cite{Faugere02}) in the future.

\begin{example}
The Gr\"obner basis of the syzygy module of Example~\ref{ex::syzygy} can be computed in 
Isabelle/HOL as
\begin{lstlisting}
value [code]
  "syzygy_basis_drlex [Vec$_0$ 0 (X * Y - Z), Vec$_0$ 0 (X * Z - Y), Vec$_0$ 0 (Y * Z - X)]"
\end{lstlisting}
which returns
\begin{lstlisting}
[
  Vec$_0$ 0 (- X * Z + Y) + Vec$_0$ 1 (X * Y - Z),
  Vec$_0$ 0 (- Y * Z + X) + Vec$_0$ 2 (X * Y - Z),
  Vec$_0$ 1 (- Y * Z + X) + Vec$_0$ 2 (X * Z - Y),
  Vec$_0$ 0 (Y - Y * Z ^ 2) + Vec$_0$ 1 (Y ^ 2 * Z - Z) + Vec$_0$ 2 (Y ^ 2 - Z ^ 2)
]
\end{lstlisting}
\const{X}, \const{Y} and \const{Z} are predefined constructors of scalar 
polynomials, representing the first three indeterminates,\footnote{More such constructors can be 
added on-the-fly when needed.} and 
$\const{Vec}_0(i,p)$ turns the scalar polynomial $p$ into a vector of polynomials by setting the 
$i$-th component to $p$ and all others to $0$.
\end{example}



\section{Conclusion}
\label{sec::Conclusion}

We hope we could convince the reader that the work described in this paper is an elegant, generic 
and executable formalization of an interesting and important mathematical theory in the realm of 
commutative algebra. Even though other formalizations of Gr\"obner bases in other proof assistants 
exist, ours is the first in Isabelle/HOL, and the first featuring the $F_4$ algorithm and Gr\"obner bases for modules. Besides, our work also gives an affirmative answer to the 
question whether multivariate polynomials \'a la \cite{Haftmann2014} can effectively be used for 
formalizing theorems and algorithms in computer algebra.

Our own contributions to multivariate polynomials make up approximately 8600 lines of proof,
and the Gr\"obner-bases related theories make up another 16700 lines of proof (3000 lines of which are about Macaulay matrices and the $F_4$ algorithm), summing up to a total of
25300 lines. Most proofs are intentionally given in a quite verbose style for better readability. The formalization effort was roughly eight person-months of full-time 
work, distributed over 
two years of part-time work. This effort is comparable to what the authors of 
other formalizations 
of Gr\"obner bases theory in other proof assistants report.


It is worth noting that all formalizations of Gr\"obner bases in existence 
restrict themselves to 
the basics of the theory. Indeed, browsing through the ample literature on 
the subject one 
quickly realizes that \emph{a lot} more properties of Gr\"obner bases, ways of 
computing them, 
generalizations, and intriguing applications could be added to the corpus of 
formal mathematics in 
the future; examples include Gr\"obner bases over coefficient rings that are no fields, \emph{elimination orders} and their applications, converting between different term orders, and non-commutative Gr\"obner bases.
We plan to contribute to this endeavor by formalizing the very 
recent approach of 
computing Gr\"obner bases by transforming \emph{Macaulay matrices} (or 
\emph{generalized Sylvester matrices}) into reduced row echelon form~\cite{WiesingerWidi15} in 
Isabelle/HOL. This approach has similarities to 
the $F_4$ algorithm but only computes the reduced row echelon form of 
\emph{one} big matrix, instead of doing this repeatedly in every iteration of a 
critical-pair/completion algorithm. Besides, formalizing said approach also 
necessitates proving upper bounds on the 
degrees of polynomials that may appear in a Gr\"obner basis, which in turn 
allows us to draw conclusions concerning the \emph{theoretical complexity} of 
algorithms for computing Gr\"obner bases. All this is ongoing work.

%


\bibliographystyle{splncs03}
\bibliography{Paper}

%
%
%

\end{document}